\documentclass[aps,prl,twocolumn]{revtex4}
\usepackage{amsfonts}
\usepackage{amsmath}
\usepackage{graphicx}
\usepackage{dsfont}
\usepackage{diagbox}
\usepackage{array}
\usepackage{amssymb}
\usepackage{bbm}
\usepackage{float}
\usepackage{subfigure}
\usepackage{url}
\usepackage{hyperref}
\usepackage{color}

\begin{document}

\title{Hybrid Berezinskii-Kosterlitz-Thouless and Ising topological phase
transition in the generalized two-dimensional XY model using tensor networks}
\author{Feng-Feng Song$^{1}$ and Guang-Ming Zhang$^{1,2}$}
\affiliation{$^{1}$State Key Laboratory of Low-Dimensional Quantum Physics and Department
of Physics, Tsinghua University, Beijing 100084, China. \\
$^{2}$Frontier Science Center for Quantum Information, Beijing 100084, China.}
\date{\today}

\begin{abstract}
In tensor network representation, the partition function of a generalized
two-dimensional XY spin model with topological integer and half-integer
vortex excitations is mapped to a tensor product of one-dimensional quantum
transfer operator, whose eigen-equation can be solved by an algorithm of
variational uniform matrix product states. Using the singularities of the
entanglement entropy, we accurately determine the complete phase diagram of
this model. Both the integer vortex-antivortex binding and half-integer
vortex-antivortex binding phases are separated from the disordered phase by
the usual Berezinskii-Kosterlitz-Thouless (BKT) transitions, while a
continuous topological phase transition exists between two different vortex
binding phases, exhibiting a logarithmic divergence of the specific heat and
exponential divergence of the spin correlation length. A new hybrid BKT and
Ising universality class of topological phase transition is thus
established. We further prove that three phase transition lines meet at a
multi-critical point, from which a deconfinement crossover line extends into
the disordered phase.
\end{abstract}

\maketitle

\textit{Introduction. -}It is well-known that topological vortices/defects
govern the critical behavior and produce rich physics in two dimensional
(2D) systems. One prominent example is the Berezinskii-Kosterlitz-Thouless
(BKT) phase transition\cite{Berezinsky_1970,Kosterlitz_1973}, which is
associated with the binding of integer vortices and antivortices in pairs
with quasi-long-range order. Such a transition cannot be characterized by
spontaneous symmetry breaking with local order parameter and constitutes the
first example of topological phase transitions beyond the Landau paradigm. A
prototype model exhibiting these fascinating features is given by the 2D
classical XY spin model, which can be realized in superfluid helium films%
\cite{Superfluid-Films} and 2D superconductor films/arrays\cite%
{Superconductor-Films1,Superconductor-Films2,Superconductor-Arrays}. When an
extra spin-nematic interaction with $\pi $ period is introduced, the
generalized XY model contains both integer vortices and half-integer
vortices along with their associated topological strings\cite%
{korshunov_1985,Lee_1985}. Although great efforts have been devoted to
establish a complete phase diagram\cite%
{Carpenter_1989,Fendley_2011,Stefan_2013,Serna_2017,Nui_2018}, there are still some
standing puzzles: what is the nature of the phase transition between the
integer vortex binding and half-integer vortex binding phases\cite%
{Lee_1985,Fendley_2011}, how does this transition line merge into two BKT
transition lines\cite{Stefan_2013} and is there a deconfinement transition
in the disordered phase\cite{Serna_2017}? The accurate determination of the
transition lines and the possible multi-critical point remain a great
challenge.

Recently, tensor network method has become a powerful theoretical tool to
characterize correlated quantum many-body phases in the thermodynamic limit%
\cite{Verstraete-Adv-Phys,Orus-Ann-Phys}. Since the partition function of a
2D classical statistical model can be represented as a tensor product of 1D
quantum transfer operator\cite{Haegeman-Verstraete2017}, the eigen-equation
of this transfer operator can be solved by the variational uniform matrix
product state (VUMPS) algorithm\cite%
{VUMPS,Fishman_2018,Verstraete-SciPostPhysLectNotes}, and all thermodynamic
properties are thus accurately obtained. In this scheme, the critical
temperature of the BKT transition in the usual 2D XY model is estimated with
high precision and the exponential divergence of the spin correlation length
is confirmed\cite{Laurens-XY}. Therefore, it becomes possible to apply the
tensor network approach to resolve those puzzles about the generalized XY
spin model, shedding new light on the nature of the topological phase
transition between two different vortex binding phases.

In this paper, we carry out such tensor network calculations for the
generalized 2D XY spin model, and accurately determine the topological phase
transitions using the singularity of the entanglement entropy associated to
the 1D quantum transfer operator. Thus the complete phase diagram of the
generalized XY model is precisely obtained. In particular, we show that a
continuous phase transition occurs between two different vortex binding
phases, where the logarithmic divergence of the specific heat of the 2D
Ising transition and the exponential divergence of the spin correlation
length of the BKT transition simultaneously exist, establishing a new
hybrid BKT and Ising universality class with two different correlation lengths. 
From the analysis of the entanglement spectrum, a deconfinement crossover 
line is found in the disordered phase, separating the region with dominant 
integer vortex excitations from the region with dominant half-integer vortex 
excitations.

\textit{Model and Method}. -The generalized XY spin model on a 2D square
lattice is given by
\begin{equation}
H=-J\sum_{\langle ij\rangle }\left[ \Delta \cos \left( \theta _{i}-\theta
_{j}\right) +(1-\Delta )\cos \left( 2\theta _{i}-2\theta _{j}\right) \right]
,  \label{Model}
\end{equation}%
where $\theta _{i}\in \lbrack 0,2\pi ]$ denotes the spin orientation at the
lattice site $i$ and the summation runs over all nearest neighbor sites. $%
\Delta =1$ corresponds to the usual XY spin model with integer vortex
excitations, while $\Delta =0$ corresponds to a purely spin-nematic model
with the invariance under $\theta _{i}\rightarrow \theta _{i}+\pi $. In the
latter, the main topological excitations are half-integer vortices and their
associated strings across which spins are antiparallel, and the so-called
half-BKT transition occurs at the exactly same critical temperature as the
usual XY model, characterized by the binding of half-integer vortices and
antivortices in pairs at low temperatures. For $0<\Delta <1$, two local
interactions strongly compete with each other, and the system undergoes a 
phase transition from the integer vortex-antivortex binding phase to the 
half-integer vortex-antivortex binding phase\cite{Lee_1985}.

\begin{figure}[tbp]
\centering
\includegraphics[width=0.45\textwidth]{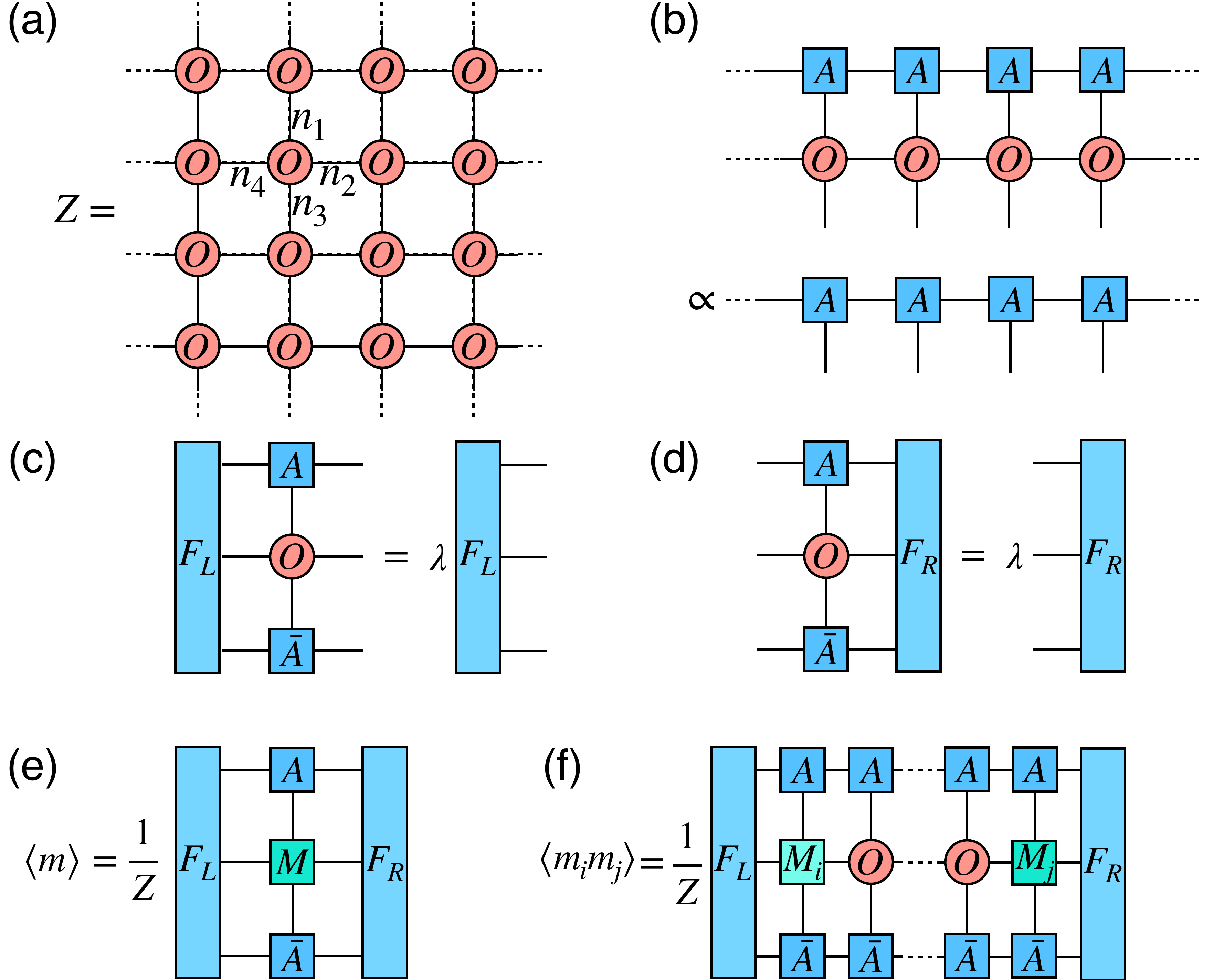}
\caption{ (a) Tensor network representation of the partition function. (b)
Eigen-equation for the fixed-point uMPS of the transfer operator $T(\protect%
\beta ,\Delta )$. (c) and (d) Eigen-equations for the left and right
fixed-point eigenvectors of the channel operator $\mathbb{T}_{O}$. (e)
Expectation of a local operator by contracting the leading eigenvectors of
the channel operators. (f) Two-point correlation function represented by
contracting a sequence of channel operators. }
\label{fig:tensor}
\end{figure}

The tensor network method starts from expressing the partition function as a
tensor network
\begin{equation}
Z=\prod_{i}\int \frac{\mathrm{d}\theta _{i}}{2\pi }\prod_{\langle ij\rangle }%
\mathrm{e}^{\beta \lbrack \Delta \cos (\theta _{i}-\theta _{j})+(1-\Delta
)\cos 2(\theta _{i}-\theta _{j})]},  \label{eq:defZ}
\end{equation}%
where the temperature $T$ is in the unit $J/k_{B}$. To find its tensor
network representation, we take a duality transformation, which changes the
phase variables into number indices on the links. Such a transformation is
obtained by the character expansion for the Boltzmann factor $e^{x\cos\theta
}=\sum_{n=-\infty }^{\infty }I_{n}(x)e^{in\theta }$, where $I_{n}(x)$ are
the modified Bessel functions of the first kind. Then the partition function
is written as
\begin{equation}
Z=\prod_{s}\int \frac{\mathrm{d}\theta _{s}}{2\pi }\prod_{l\in \mathcal{L}%
}\sum_{n_{l}}a_{n_{l}}(\beta ,\Delta )\mathrm{e}^{in_{l}\left( \theta
_{s_{i}}-\theta _{s_{j}}\right) },  \label{eq:3}
\end{equation}%
where $a_{n}(\beta ,\Delta )=\sum_{m=-\infty }^{+\infty }I_{n-2m}(\beta
\Delta )I_{m}(\beta (1-\Delta ))$, $l$ runs over all the links and $s$
labels all the lattice sites. By integrating out the physical degrees of
freedom $\theta $, the partition function is represented as a tensor network
\begin{equation}
Z=\mathrm{tTr}\,\prod_{s}O_{n_{1},n_{2}}^{n_{3},n_{4}}\left( s\right) ,
\end{equation}%
where $\mathrm{tTr}$ denotes the tensor contraction and each local tensor is
given by $O_{n_{1},n_{2}}^{n_{3},n_{4}}=\left(
\prod_{i=1}^{4}a_{n_{i}}(\beta ,\Delta )\right)
^{1/2}\delta_{n_{1}+n_{2}}^{n_{3}+n_{4}}$. The partition function is shown
in Fig.~\ref{fig:tensor} (a). The continuous symmetry of the model has been
encoded in the tensor network representation as the constraint of vanishing
lattice divergence: $O_{n_{1},n_{2}}^{n_{3},n_{4}}\neq 0$ only if 
$n_{1}+n_{2}-n_{3}-n_{4}=0$, i.e., the conservation law of $U(1)$ charges. 
Since the expansion coefficients decrease exponentially
with increasing $n$, an appropriate truncation can be performed on the
virtual legs of the tensor $O$ without loss of accuracy. Actually, the dual 
index $n$ plays the essential role in describing the Ising 
physics when interpreted as the curl of the height variable $h_a$ on the dual 
lattice, where the corresponding dual Ising spin is defined by\cite{Serna_2017}
$\mu_a=(-1)^{h_a}$.

The fundamental object in the tensor-network partition function is the
row-to-row quantum transfer operator,
\begin{equation*}
\widehat{T}(\beta ,\Delta )=\sum_{\ldots s,p,q\ldots }\mathrm{Tr}\,\left(
\cdots O_{n_1}^{n_3}(s)O_{n_1}^{n_3}(p)O_{n_1}^{n_3}(q)\cdots \right) ,
\label{eq:TM}
\end{equation*}%
where $s,p,q\ldots $ refer to the lattice sites in a row. This operator
plays the same role as the matrix product operator (MPO) for 1D quantum spin
chains\cite{Haegeman-Verstraete2017}, whose model Hamiltonian can be
formally expressed as
\begin{equation}
\hat{H}_{1d}=-\frac{1}{\beta}\log\widehat{T}(\beta ,\Delta ).
\end{equation}
Then the partition function is determined by the dominant eigenvalues of $%
\widehat{T}(\beta ,\Delta )$. For a translation-invariant MPO, the leading
eigenvector as the fixed-point of $\widehat{T}(\beta ,\Delta )$ can be
represented by a uniform matrix product states (uMPS)\cite{VUMPS}
\begin{equation}
|\Psi (A)\rangle =\sum_{\ldots s,p,q\ldots }\mathrm{Tr}\,\left( \cdots
A^{n_{s}}A^{n_{p}}A^{n_{q}}\cdots \right) ,  \label{eq:MPS}
\end{equation}%
where $A_{\alpha \beta }^{n_{s}}$ is a three-leg tensor with bond dimension $%
D$, controlling the accuracy of this approximation. Since the transfer
operator $\widehat{T}(\beta ,\Delta )$ is Hermitian, the eigen-equation as
shown in Fig.~\ref{fig:tensor} (b)
\begin{equation}
\widehat{T}(\beta ,\Delta )|\Psi (A)\rangle =\Lambda _{\max }|\Psi
(A)\rangle ,  \label{eq:fixpoint}
\end{equation}%
can be transformed as an optimization problem
\begin{equation*}
\max_{A}\langle \Psi (A)|\widehat{T}(\beta ,\Delta )|\Psi (A)\rangle
/\langle \Psi (A)|\Psi (A)\rangle ,
\end{equation*}%
for the local tensor $A$. To solve this optimization problem, we apply the
VUMPS algorithm\cite{VUMPS,Fishman_2018,Verstraete-SciPostPhysLectNotes},
which provides an efficient variational scheme to approximate the largest
eigenvector $|\Psi (A)\rangle $. Then various physical quantities can be
estimated from the fixed-point uMPS.

For the entanglement entropy, we perform a bipartition on $|\Psi (A)\rangle $
via Schmidt decomposition
\begin{equation}
|\Psi (A)\rangle =\sum_{\alpha ,\beta =1}^{D} s_{\alpha }\delta _{\alpha
,\beta }|\Psi _{\alpha }^{-\infty ,n}\rangle |\Psi _{\beta }^{n+1,\infty
}\rangle .  \label{eq:svd}
\end{equation}%
where $s_{\alpha }$ are the singular values. The entanglement entropy $S_{E}$
is given by
\begin{equation}
S_{E}=-\sum_{\alpha =1}^{D} s_{\alpha }^{2}\ln s_{\alpha }^{2}  \label{eq:ee}
\end{equation}%
in the same way as the quantum entanglement measure for a many-body quantum
state\cite{Vidal_2003}. Moreover, the entanglement spectrum\cite%
{Haldane_2008} can be defined by $\varepsilon_{n}=-\log s_{n}^2$ to yield
more information on the fixed-point uMPS.

Moreover, the evaluations of a local observable $m(\theta _{i})$ can be
represented as
\begin{equation}
\left\langle m\left( \theta _{i}\right) \right\rangle =\frac{1}{Z}%
\prod_{j}\int \frac{\mathrm{d}\theta _{j}}{2\pi }\mathrm{e}^{-\beta E\left(
\left\{ \theta _{j}\right\} \right) }m\left( \theta _{i}\right) ,
\label{eq:single}
\end{equation}%
where $E(\{\theta _{j}\})$ is the energy under a given configuration $%
\{\theta _{j}\}$. Compared to the partition function \eqref{eq:defZ}, the $%
m(\theta _{i})$ on the right hand side of Eq.\eqref{eq:single} simply
changes the tensor $O_{n_{1},n_{2}}^{n_{3},n_{4}}(i)$ into an impurity local
tensor
\begin{equation*}
M_{n_{1},n_{2}}^{n_{3},n_{4}}=\left( \prod_{i=1}^{4}a_{n_{i}}(\beta ,\Delta
)\right) ^{1/2}\int \frac{\mathrm{d}\theta }{2\pi }\mathrm{e}^{i\theta
\left( n_{1}+n_{2}-n_{3}-n_{4}\right) }m(\theta ).
\end{equation*}%
Using the uMPS fixed-point, the contraction of the tensor network of $%
\left\langle m\left( \theta \right) \right\rangle $ is reduced to a trace of
an infinite sequence of channel operators
\begin{equation}
\left\langle m\left( \theta \right) \right\rangle =\mathrm{Tr}\left( \cdots
\mathbb{T}_{O}\mathbb{T}_{O}\mathbb{T}_{M}\mathbb{T}_{O}\mathbb{T}_{O}\cdots
\right) ,  \label{eq:channelsingle}
\end{equation}%
where the channel operator is defined by
\begin{equation}
\mathbb{T}_{X}=\sum_{i,j}\bar{A}^{i}\otimes X^{i,j}\otimes A^{j}.
\label{eq:channel}
\end{equation}%
In the same fashion, the contraction of channel operators is determined by
the leading eigenvectors $\langle F_{L}|$ and $|F_{R}\rangle $ of $\mathbb{T}%
_{O}$ as shown in Fig.~\ref{fig:tensor} (c) and (d). Thus the expectation
value of a local observable can be obtained by contraction of the leading
eigenvectors $\langle F_{L}|$ and $|F_{R}\rangle $ with $\mathbb{T}_{M}$,
\begin{equation}
\left\langle m\left( \theta \right) \right\rangle =\langle F_{L}|\mathbb{T}%
_{M}|F_{R}\rangle ,  \label{eq:tensorlsingle}
\end{equation}%
as shown in Fig.~\ref{fig:tensor} (e). Moreover, the correlation functions
between two local observable $m(\theta _{i})$ and $m(\theta _{j})$ defined
by $G(r)=\langle m(\theta _{i})m(\theta _{j})\rangle $ can also be
represented as a contraction of a tensor network with two local impurity
tensors $M_{i}$ and $M_{j}$. As displayed in Fig.~\ref{fig:tensor} (f), with
the help of fixed-points of the channel operator, we can easily derive
\begin{equation}
G(r)=\langle F_{L}|\mathbb{T}_{M_{i}}\underbrace{\mathbb{T}_{O}\cdots
\mathbb{T}_{O}}_{r-1}\mathbb{T}_{M_{j}}|F_{R}\rangle .  \label{eq:tensorlcl}
\end{equation}

\begin{figure}[tbp]
\centering
\includegraphics[width=0.45\textwidth]{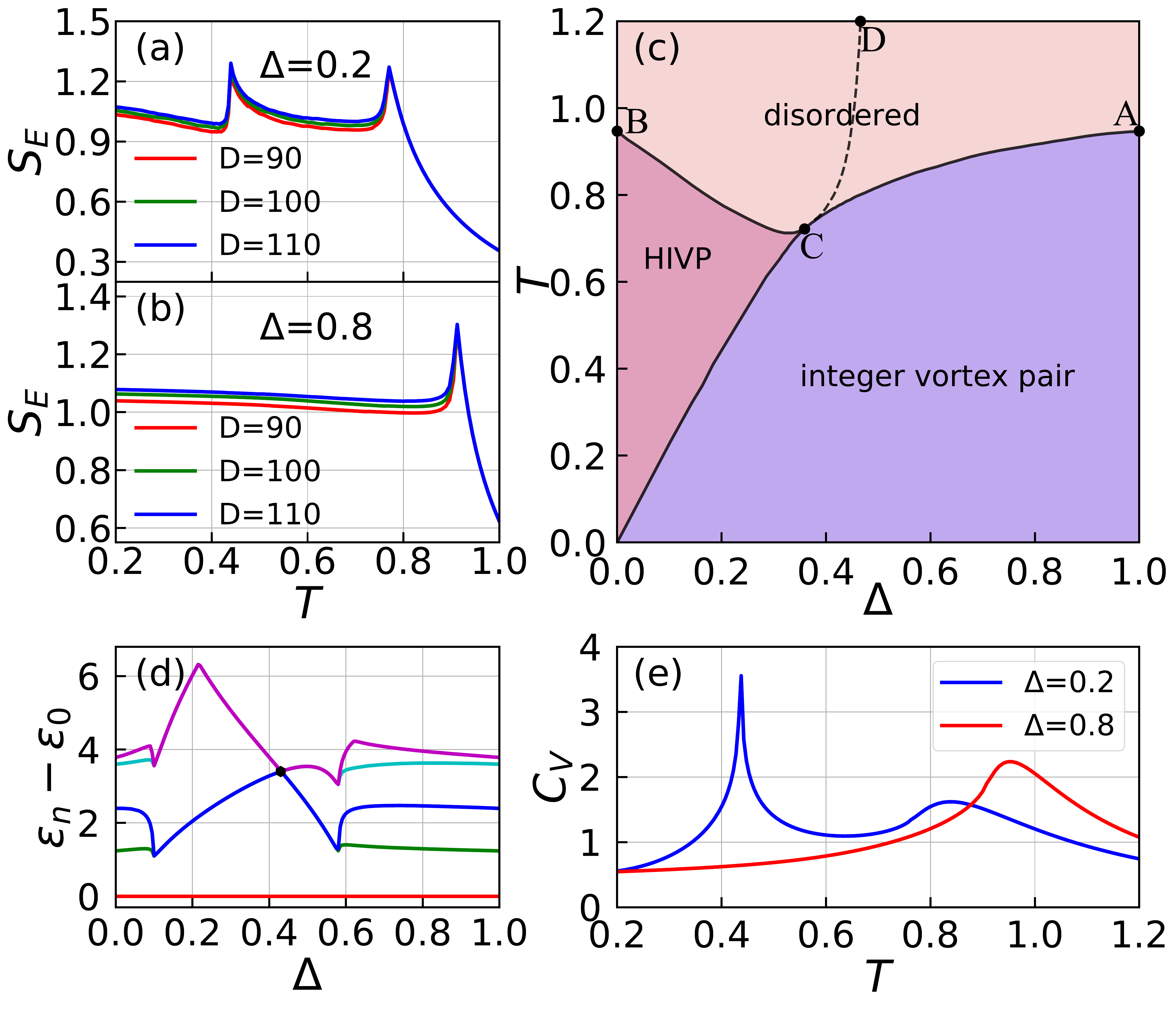}
\caption{ (a) and (b) The entanglement entropy as a function of temperature
for $\Delta =0.2$ and $\Delta =0.8$. (c) The complete phase diagram of the
generalized 2D XY model. $AC$ and $BC$ are two BKT transition lines, and
there is a special continuous transition line from the origin to a
multi-critical point C. HIVP is an abbreviation for the half-integer vortex
pair. (d) The five lowest entanglement levels of the fixed-point uMPS under $%
T=0.85$. There is a four-fold degenerate crossing point (black dot)
corresponding to the deconfinement crossover in the disordered phase. (e)
The specific heat as a function of temperature for $\Delta =0.2$ and $\Delta
=0.8$. }
\label{fig:phasediagram}
\end{figure}

\textit{Numerical results}. -In the tensor network framework, the
entanglement entropy of the fixed-point uMPS for the 1D quantum transfer
operator exhibits singularity, which can be used to accurately determine all
possible phase transitions. First, we estimate the critical temperature of
the BKT transition for the usual XY model ($\Delta =1$) as $T_{\text{BKT}%
}\simeq 0.8933$, sufficiently close to the result from high-temperature
expansion with high precision\cite{Arisue_2009,Hsieh_2013}. Then we
calculate the entanglement entropy for a general value of $0<\Delta <1$. In
Fig.~\ref{fig:phasediagram}(a) and (b), the entanglement entropy $S_{E}$
develops two sharp peaks at $T_{c,1}\simeq 0.44$ and $T_{c,2}\simeq 0.77$
for a small value $\Delta =0.2$, and one sharp peak at $T_{c}\simeq 0.91$
for a large value $\Delta =0.8$. The peak positions are almost unchanged
with the bond dimensions $D=90,100,110$. So we can accurately locate the
phase boundaries, and the complete phase diagram is thus derived as
displayed in Fig.~\ref{fig:phasediagram}(c). We find two BKT phase
transition lines $AC$ and $BC$ corresponding to the integer
vortex-antivortex binding and the half-integer vortex-antivortex binding
transitions, respectively. More importantly, there is another topological
phase transition line between the integer vortex-antivortex binding phase
and the half-integer vortex-antivortex binding phase, and three phase
transition lines merge at a multi-critical point $C$ ($\Delta \simeq 0.330$,
$T\simeq 0.705$).

Furthermore the entanglement spectrum of the fixed-point uMPS for the 1D
quantum transfer operator exhibits some intriguing features. In Fig.~\ref%
{fig:phasediagram} (d), the five lowest entanglement levels are displayed
for a given temperature $T=0.85$, where varying $\Delta $ goes through all
three phases. The lowest entanglement level is always non-degenerate.
Exactly at the two critical BKT transitions $\Delta _{c,1}\simeq 0.10$ and $%
\Delta _{c,2}\simeq 0.58$, the first and second excited entanglement levels
become degenerate when entering the disordered phase. Inside the disordered
phase, we find an unusual feature that the two-fold degenerate excited
levels have a further crossing to form a four-fold degenerate point at $%
\Delta ^{\ast }\simeq 0.43$. With a careful numerical analysis, we find that
this four-fold degenerate point just starts from the multi-critical point $C$
and extends into the disordered phase. In the phase diagram Fig.~\ref%
{fig:phasediagram} (c), these highly degenerate points are plotted in a
dotted line, which approaches $\Delta =0.5$ asymptotically as increasing the
temperature. According to the discussion from the dual height model\cite%
{Serna_2017}, we identify this special line as the deconfinement crossover
line separating the region with dominant integer vortex excitations ($%
0.5<\Delta<1$) from the half-integer vortex dominant region ($0<\Delta<0.5$).

To further understand the topological phase transition between two different
vortex binding phases, we calculate the specific heat, which can be
represented in the tensor-network language based on two nearest neighbor
impurity tensors. Using the contraction in \eqref{eq:tensorlcl}, the
internal energy per site can be calculated as
\begin{equation*}
u=-\Delta \langle \cos \left( \theta _{i}-\theta _{i+1}\right) \rangle
-(1-\Delta )\langle \cos \left( 2\theta _{i}-2\theta _{i+1}\right) \rangle ,
\end{equation*}%
and the specific heat per site is given by $C_{V}=\partial u/\partial T$.
For $\Delta =0.2$, all three phases can be reached as varying the
temperature, while for $\Delta =0.8$ there are only two phases. As shown in
Fig.~\ref{fig:phasediagram}(e), the specific heat shows a bump around two
BKT type transitions but a logarithmic divergence at the transition between
two different vortex binding phases. Such a logarithmic specific heat is
usually regarded as a characteristic feature of the 2D Ising transition\cite%
{korshunov_1985,Lee_1985,Carpenter_1989,Fendley_2011,Stefan_2013,Serna_2017}.

\begin{figure}[tbp]
\centering
\includegraphics[width=0.45\textwidth]{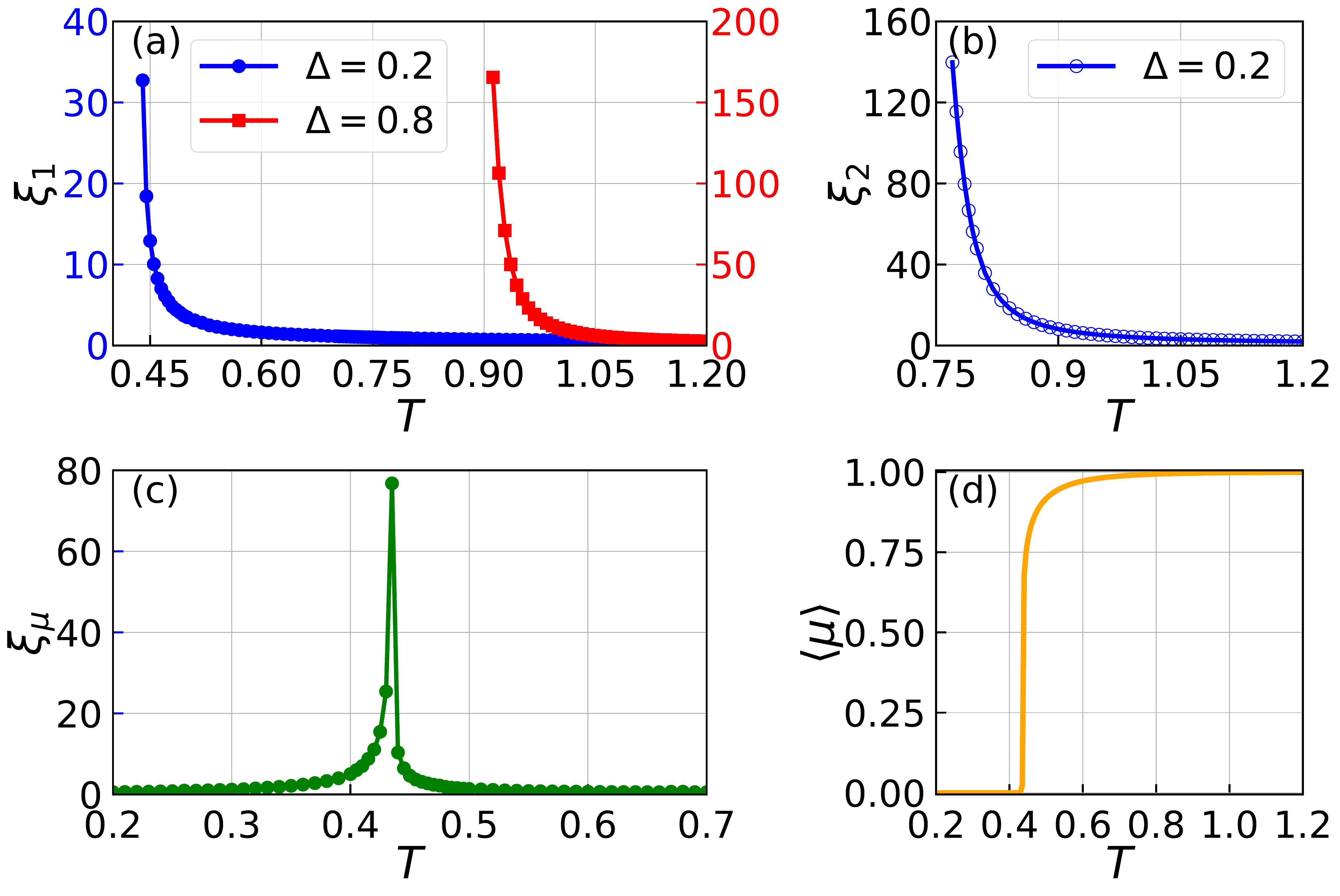}
\caption{ (a) and (b) Correlation lengths derived from the spin and nematic
spin correlation functions with bond dimension $D=110$. (c) The dual Ising
correlation length extracted from the dual Ising spin correlator along $%
\Delta=0.2$. (d) Expectation value of the dual Ising spins along the line $%
\Delta=0.2$ in the phase diagram.}
\label{fig:correlationlength}
\end{figure}

Actually the nature of this phase transition between two different vortex
binding phases can also be revealed by studying the following correlation
functions
\begin{equation*}
G_{1}(r)=\langle \cos (\theta _{i}-\theta _{i+r})\rangle ,\quad
G_{2}(r)=\langle \cos (2\theta _{i}-2\theta _{i+r})\rangle ,
\end{equation*}%
which describe the correlations of the spins and nematic spins,
respectively. Within the tensor network framework, these two correlation
functions can be calculated readily using local impurity tensors of $e^{\pm
i\theta _{i}}$ and $e^{\pm i2\theta _{i}}$. It has been known that $%
G_{1}\left( r\right) $ decays algebraically \textit{only} in the integer
vortex-antivortex binding phase, while $G_{2}\left( r\right) $ exhibits
algebraic decay \textit{both} in the half-integer vortex-antivortex binding
and integer vortex-antivortex binding phases\cite{Lee_1985,Carpenter_1989}.
When $G_{1}(r)$ and $G_{2}(r)$ decay exponentially, two different
characteristic lengths $\xi _{1}$ and $\xi _{2}$ are defined by\cite{Nui_2018}
\begin{equation}
G_{1}(r)\sim \exp (-r/\xi _{1}),\quad G_{2}(r)\sim \exp (-r/\xi _{2}).
\end{equation}%
In particular, $G_{1}\left( r\right) $ decays exponentially in the
half-integer vortex-antivortex binding phase, so we can study the critical
behavior of the correlation length $\xi _{1}$ when approaching the phase
transition. In Fig.~\ref{fig:correlationlength} (a), we show the correlation
length $\xi _{1}$ as a function of temperature for $\Delta =0.2$ with a
sharp divergence above the critical point $T_{c}\simeq 0.44$, as well as $%
\Delta =0.8$ with a sharp divergence above the BKT transition point $%
T_{c}\simeq 0.91$. Similarly, the correlation length $\xi _{2}$ for $\Delta
=0.2$ as displayed in Fig.~\ref{fig:correlationlength} (b) has a divergence
above the half-BKT transition point $T_{c}\simeq 0.77$. When approaching the
critical points from the high-temperature side, all correlation lengths are
found to be well-fitted by an exponentially divergent form
\begin{equation}
\xi (T)\propto \exp (\frac{b}{\sqrt{T-T_{C}}}),T\rightarrow T_{C}^{+}
\label{eq:KT_cl}
\end{equation}%
where $b$ is a non-universal positive constant, which is a characteristic
property of the BKT transition\cite{Kosterlitz_1974}.

Moreover, the two-point correlator characterizing the Ising strings is defined 
by
\begin{eqnarray}
C(r)&=&\langle \mu(i)\mu(i+r) \rangle \notag \\
&=&\langle (-1)^{n_i}(-1)^{n_{i+1}}\cdots(-1)^{n_{i+r-1}}\rangle.
\end{eqnarray}
So the correlation function of the dual Ising spin variables is transformed
into an expectation of a string of impurity variables $(-1)^{n_i}$. On both
sides of the critical point $T_c\simeq0.44$, we find $C(r)$ decays
exponentially, and the corresponding Ising correlation length $\xi_{\mu}$ is
extracted and displayed in Fig.~\ref{fig:correlationlength} (c). When
approaching $T_c$ from either side, $\xi_{\mu}$ diverges as $1/|T-T_c|$ in
the same way of the 2D Ising transition. In Fig.~\ref{fig:correlationlength}(d) 
we also show the dual Ising order parameter and $\langle\mu\rangle>0$ 
indicates the proliferation of Ising domain walls. All these results strongly 
suggest that the transition from the half-integer vortex-antivortex binding 
phase to the integer vortex-antivortex binding phase is a hybrid BKT and Ising
transition. Two different correlation lengths are needed to characterize this 
new universality class of topological phase transition.

\begin{figure}[tbp]
\centering
\includegraphics[width=0.45\textwidth]{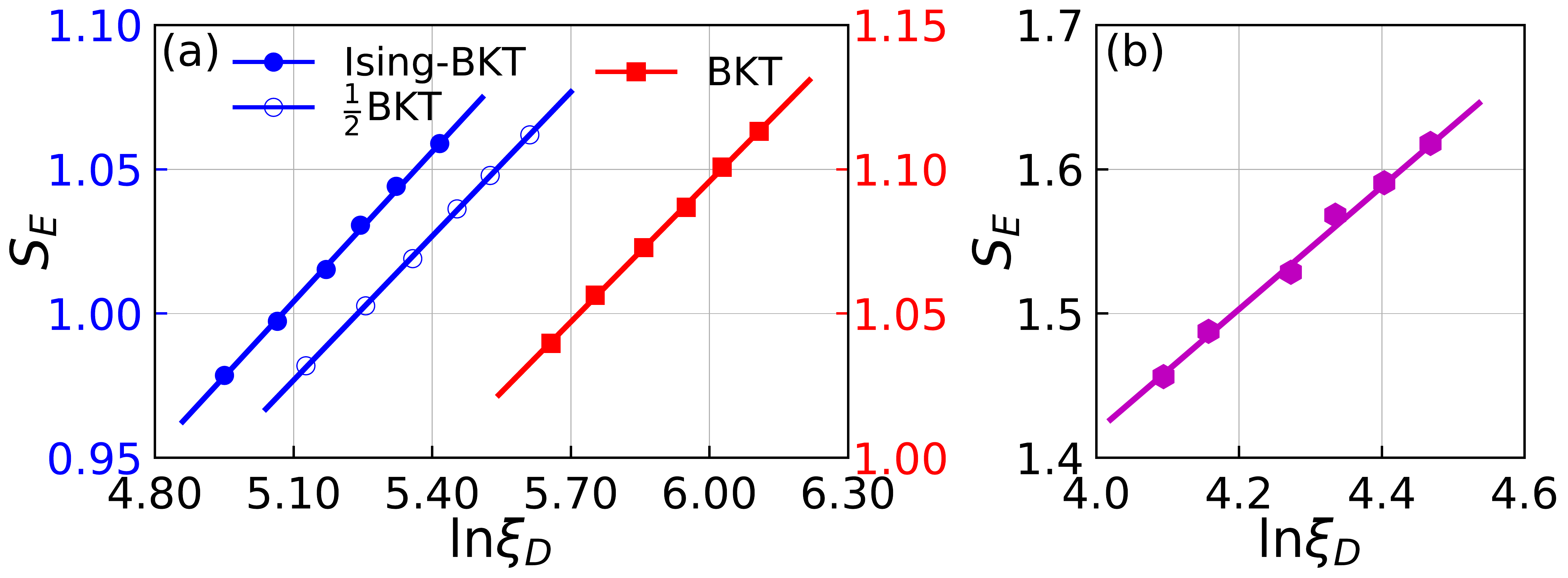}
\caption{ (a) Central charges extracted from a linear fit of the
entanglement entropy with increasing the bond dimensions from $110$ to $160$%
: $c\simeq 0.992$ for the Ising-BKT transition, $c\simeq 0.999$ for the BKT
transition, and $c\simeq 0.996$ for the half-BKT transition. (b) The central
charge $c\simeq 2.492$ is extracted from the correlation length close to the
multi-critical point $C$.}
\label{fig:centralcharge}
\end{figure}

Furthermore, we can extract the central charges for those critical field
theories by numerically fitting the entanglement entropy with the
correlation lengths obtained under different bond dimensions\cite%
{PhysRevLett.102.255701}
\begin{equation}
S_{E}\propto \frac{c}{6}\ln \xi _{D},  \label{eq:centralcharge}
\end{equation}
where $\xi_D$ is the largest correlation length with the different bond 
dimensions. In Fig.~\ref{fig:centralcharge} (a), the numerical fitting clearly 
shows the central charges equal to $c=1$ for all three transitions, precisely 
the value of the BKT-type transition. More importantly, close to the 
multi-critical point $C$, we can also deduce the leading correlation length
from the fixed-point uMPS from the disorder phase. In Fig.~\ref{fig:centralcharge}(b), 
by fitting the correlation lengths determines the corresponding central charge
as $c\simeq5/2$. There is a possible way to explain this result. Since two 
different correlation lengths are involved in the
Ising-BKT transition, the corresponding critical field theory should be 
characterized by a 2D conformal field theory with a central charge $c=1+1/2$, 
and this Ising-BKT transition line further merges with the BKT transition line 
$ACB$ at the multi-critical point $C$, leading to the central charge $c=1+3/2$.

\textit{Conclusion and Outlook}. -We have employed the tensor network method
to solve the generalized two-dimensional classical XY spin model with
topological integer and half-integer vortex excitations as well as the
string excitations. Using the singularity of the entanglement entropy, we
have accurately determined the phase diagram of this model. In particular, a
new hybrid BKT and Ising phase transition has been established between the
integer vortex-antivortex binding phase and the half-integer
vortex-antivortex binding phase. To describe this new phase transition, two
distinctly characteristic correlation lengths are necessary, the corresponding 
critical field theory may be characterized by a 2D conformal field theory with 
the $Z_{2}\times U(1)$ symmetry.

In recent years, a large amount of investigations have focused on a gas of 
attractive bosons, which can form two distinct superfluid phases: an atomic 
superfluid of bosons and a molecular superfluid of boson pairs\cite%
{Stoof_2004,Weichman_2004,Stefan_2012}. The former corresponds to the
integer vortex-antivortex binding phase and the latter is the half-integer
vortex-antivortex binding phase. The established hybrid topological phase
transition may be experimentally realized in these physical systems.

\begin{acknowledgments}
\textbf{Acknowledgments}
\end{acknowledgments}

The research is supported by the National Key Research and Development
Program of MOST of China (2017YFA0302902).

\bibliography{reference}

\end{document}